\documentclass[a4paper]{article}

\usepackage{INTERSPEECH2020}

\usepackage[T1]{fontenc}    % use 8-bit T1 fonts
\usepackage{hyperref}       % hyperlinks
\usepackage{url}            % simple URL typesetting
\usepackage{booktabs}       % professional-quality tables
\usepackage{amsfonts}       % blackboard math symbols
\usepackage{nicefrac}       % compact symbols for 1/2, etc.
\usepackage{microtype}      % microtypography
\usepackage{color}
\usepackage{algorithm,algpseudocode}
\usepackage{subcaption}

\usepackage{multirow}
\usepackage{amssymb}
\usepackage{amssymb}% http://ctan.org/pkg/amssymb
\usepackage{pifont}
\newcommand{\cmark}{\ding{51}}%
\newcommand{\xmark}{\ding{55}}%
\usepackage{color}
\usepackage{algorithm}
\usepackage{cite}
\usepackage{algorithmicx}
\usepackage[hang,flushmargin]{footmisc}

\title{Audio-visual Multi-channel Recognition of Overlapped Speech}
\name{Jianwei Yu$^{1,2}$\thanks{This work was done while the author was an intern at Tencent AI Lab.},  Bo Wu$^2$, Rongzhi Gu$^2$, Shi-Xiong Zhang$^2$, Lianwu Chen$^2$, Yong Xu$^2$, Meng Yu$^2$, \\ Dan Su$^2$, Dong Yu$^2$, Xunying Liu$^1$, Helen Meng$^1$}
\address{
  $^1$The Chinese University of Hong Kong, \ \ \ \
  $^2$Tencent AI Lab}
\email{\{jwyu,xyliu,hmmmeng\}@se.cuhk.edu.hk, \{lambowu,auszhang,dyu\}@tencent.com}
\begin{document}
\maketitle
\begin{abstract}
% Automatic speech recognition (ASR) of overlapped speech remains a highly challenging task to date. 
% To this end, many state-of-the-art ASR systems have adopted multi-channel integration techniques as speech separation front-ends to enhance the target speaker's signal. 
% % The overall system can be divided into two parts: speech separation and speech recognition.
% Motivated by the fact that visual modality is invariant to acoustic signal corruption, this paper proposed an audio-visual multi-channel overlapped speech recognition system featuring tightly integrated separation back-end and recognition font-end. 
% A series of audio-visual multi-channel speech separation networks based on \textit{TF masking}, \textit{mask-based MVDR} and \textit{filter\&sum} approaches were developed. 
% To reduce the error cost mismatch between the separation front-end and recognition back-end, the two components are jointly finetuned in the system development using connectionist temporal classiﬁcation (CTC) loss function or multi-task based interpolation of CTC and scale-invariant signal to noise ratio (Si-SNR) error costs. 
% Experiments suggest that the proposed audio-visual multi-channel recognition system outperforms the baseline audio-only multi-channel ASR systems by up to 9.27\% (36.52\% relative) and 22.22\% (56.87\% relative) absolute word error rate (WER) reduction on overlapped speech constructed using either simulation or replaying of the lipreading sentence 2 (LRS2) dataset respectively.
Automatic speech recognition (ASR) of overlapped speech remains a highly challenging task to date. 
To this end, multi-channel microphone array data are widely used in state-of-the-art ASR systems. 
Motivated by the invariance of visual modality to acoustic signal corruption, this paper presents an audio-visual multi-channel overlapped speech recognition system featuring tightly integrated separation front-end and recognition back-end. 
A series of audio-visual multi-channel speech separation front-end components based on \textit{TF masking}, \textit{filter\&sum} and \textit{mask-based MVDR} beamforming approaches were developed. 
To reduce the error cost mismatch between the separation and recognition components, they were jointly fine-tuned using the connectionist temporal classification (CTC) loss function, or a multi-task criterion interpolation with scale-invariant signal to noise ratio (Si-SNR) error cost. 
Experiments suggest that the proposed multi-channel AVSR system outperforms the baseline audio-only ASR system by up to 6.81\% (26.83\% relative) and 22.22\% (56.87\% relative) absolute word error rate (WER) reduction on overlapped speech constructed using either simulation or replaying of the lipreading sentence 2 (LRS2) dataset respectively.

\end{abstract}

\section{Introduction}

% Paragraph 1.
% Despite the rapid progress in automatic speech recognition (ASR) in the past few decades, recognizing overlapped speech remains a highly challenging task. 
% The presence of interfering speakers creates a large mismatch against clean speech and leads to significant performance degradation in current ASR systems.
% Beamforming techniques integrate sensor data for multiple array channels - they "listen" in the target speaker's direction while attenuating the effects of noise distortions and interfering speakers.
% The desired speaker signal is thereby enhanced, while noise distortions and interfering speakers are attenuated.
% Many state-of-the-art ASR systems have used mircophone arrays, often using a traditional speech enhancement based approach.
% This split the overall system into two parts: speech separation and recognition components.
% The separation components are often realized using traditional beamforming techniques represented by either time domain delay and sum \cite{beamformit,DSB} or frequency domain  minimum variance distortionless response (MVDR) \cite{MVDR} approaches.
% The former uses generalized cross entropy with phase transformation and vitebi search to compute the optimal delay and channel weights. Whereas, the later maximize the signal to noise ration (SNR). 

Despite the rapid progress in automatic speech recognition (ASR) in the past few decades, recognizing overlapped speech remains a highly challenging task. 
The presence of interfering speakers creates a large mismatch against clean speech, which leads to a significant performance degradation in current ASR systems. 
To this end, acoustic beamforming techniques integrating sensor data from multiple array channels are usually adopted. These approaches "listen" in the speaker's direction while attenuate the effects of noise distortions and interfering speakers. The desired speaker signal is thereby enhanced. 
Many state-of-the-art ASR systems have used microphone arrays, often following a traditional speech enhancement based approach.
% Many state-of-the-art ASR systems have used mircophone arrays via traditional speech enhancement based approaches.
% Many state-of-the-art ASR systems have adopted traditional speech enhancement based approaches using microphone arrays.
This splits the overall system into two parts: speech separation and speech recognition. 
The separation components are often realized using conventional beamforming techniques represented either by time domain delay and sum \cite{DSB, beamformit} or frequency domain  minimum variance distortionless response (MVDR) \cite{MVDR2001, MVDR} and generalized eigenvalue (GEV) \cite{GEV2007} approaches. 
The former uses generalized cross entropy with phase transformation and Viterbi search to compute the optimal delay and channel weights, while the latter maximizes the signal to noise ratio (SNR). 

% Paragraph 2
% The success of deep learning based speech technologies allow microphone array channel integration methods to involve into wide range of neural network based designs. 
% Such methods can be classified into three categories. 
% The first category, \textit{T-F masking} approaches estimate time-frequency (T-F) masks to facilitate separation from other interfering speakers.  
% The second category,  
% The third category, \textit{filter\&sum} approaches estimate the beamforming filter parameters applies the resulting filter for channel integration to produce the separated output.    
% The neural network based beamforming methods allow a tight integration with recognition system back-end to be more conveniently implemented. The use of microphone array based multi-channel inputs can greatly improve the performance of overlapped speech recognition to date. However, the performance gap against non-overlapped speech remains significant.

The success of deep learning based speech technologies allows microphone array channel integration methods to evolve into a wide range of neural network (NN) based designs. 
These methods can be roughly classified into three categories, i.e. \textit{TF masking}, \textit{filter\&sum} and \textit{mask-based MVDR} or \textit{GEV}. 
The neural network (NN) based \textit{TF masking} approaches \cite{WJTVF, MULTIBAND} predict spectral time-frequency (TF) masks that specify whether a particular TF bin is dominated by the target speaker or interfering sources to facilitate speech separation.
The neural \textit{filter\&sum} approaches directly estimate the beamforming filter parameters in either time domain \cite{Taramulti, FASNET1, FASNET2} or frequency domain \cite{DEEPBEAMFORMER} before applying these to channel integration to produce the separated output.
The more complicated \textit{mask-based MVDR} \cite{IMPMVDR, Angle, ROM, MULTIFAROVER, MIMO2019, MIMO2020} and related \textit{mask-based GEV} \cite{GEV2016, Beamnet}  approaches compute the power spectral density (PSD) matrices for the target and overlapping speakers using spectral TF masks to obtain the beamforming parameters.  
% The \textit{mask-based MVDR} approaches \cite{IMPMVDR, Angle, ROM, MULTIFAROVER, MIMO2019, MIMO2020} adopt the TF mask to estimate the target and interfering speakers' covariance matrices (SCMs) required by MVDR beamformer to obtain the beamforming filters.
% The \textit{mask-based MVDR} approaches \cite{IMPMVDR, Angle, ROM, MULTIFAROVER, MIMO2019, MIMO2020} adopt the TF mask to estimate the target and interfering speakers' covariance matrices (SCMs) to obtain the beamforming filters via MVDR solutions.
% The \textit{filter\&sum} approaches directly estimate the beamforming filters in either time domain \cite{Taramulti, FASNET1, FASNET2} or frequency domain \cite{DEEPBEAMFORMER}, and then apply the filters to channel integration to produce the separated output.
The NN based beamforming methods allow tight integration with the recognition back-end to be more conveniently implemented \cite{DEEPBEAMFORMER, Beamnet, CRM, MIMO2019, MIMO2020}. The use of microphone array based multi-channel inputs can greatly improve the performance of overlapped speech recognition. However, the performance gap between overlapped and non-overlapped speech remains large to date.

% Paragraph 3
% Visual information is inherently invariant to acoustic signal corruption. Therefore, visual modality can be used to improve the overlapped speech recognition performance. Previous research has successfully Incorporated visual modality into single channel overlapped speech separation \cite{AVSE1, AVSEWJ} and recognition \cite{AVSRJIAN, AVSR2}. An audio-visual multi-channel speech separation system was recently proposed in \cite{gu2020multi}. However, there has been very limited previous research on audio-visuak multi-channel overlapped speech recognition.

Human speech perception is bi-modal in nature\cite{zhou2019talking}. 
The visual information is inherently invariant to acoustic signal corruption.
Therefore, the visual modality can be used to improve the recognition performance on overlapped speech.
% Therefore, the visual modality can be used to improve the overlapped speech recognition performance. 
Previous research has successfully incorporated the visual modality into single-channel overlapped speech separation \cite{AVSE1, AVSEWJ, MYLIP} and recognition \cite{AVSR2016, AVSRJIAN, AVSR2, liu2019exploiting,liu2020exploiting}. 
Recently, audio-visual multi-channel systems designed for speech separation have been proposed in \cite{gu2020multi, tan2019AV}.
% An audio-visual multi-channel speech separation system has been recently proposed in \cite{gu2020multi}. 
However, there has been very limited previous research on audio-visual multi-channel recognition of overlapped speech.

% Human speech perception is biomodal in nature. 
% Considering the fact that visual modality is inherently invariant to acoustic signal corruption, leveraging visual information to improve the overlapped speech recognition performance is a promising alternative.
% Previous research has successfully incorporated visual modality into single channel overlapped speech separation \cite{AVSE1, AVSEWJ} and recognition \cite{AVSRJIAN, AVSR2}. 
% An audio-visual multi-channel speech separation system was recently proposed in \cite{gu2020multi}. 
% However, there has been very limited previous research on audio-visual multi-channel recognition of overlapped speech.

% Paragraph 5
In this paper, we proposed an audio-visual multi-channel overlapped speech recognition system featuring tightly integrated separation front-end and recognition back-end. 
First, a series of audio-visual multi-channel speech separation networks based on \textit{TF masking}, \textit{filter\&sum} and \textit{mask-based MVDR} approaches were developed respectively. 
Second, in order to reduce the error cost mismatch between the separation and recognition components, the two components are jointly fine-tuned using the CTC loss function, or a multi-task criterion interpolation with Si-SNR error cost. 
% Second, in order to reduce the error cost mismatch between the separation front-end and recognition back-end, these two components are jointly finetuned in the system development using CTC cost function or multi-task based interpolation between CTC and Si-SNR error costs. 
Experiments suggest that the proposed audio-visual multi-channel recognition system outperforms the baseline audio-only multi-channel ASR systems by up to 6.81\% (26.83\% relative) and 22.22\% (56.87\% relative) absolute WER reduction on overlapped speech constructed using either simulation or replaying of the LRS2 dataset respectively.
% {\color{red}{The performance gaps between simulated and replaying overlapped speech against non-overlapped speech are reduced by 9.27\% (92.24\% relative) and 22.22\% (93.60\% relative) absolute from a WER of 10.05\% and 23.74\% obtained using the audio-only multi-channel ASR baseline system to 0.78\% and 1.52\% produced by the proposed audio-visual multi-channel system.}}{\color{blue}{(I think we can remove this sentence.)}}
To the best of our knowledge, this paper is among the first to use audio-visual multi-channel integration for the overlapped speech recognition.

The rest of the paper is organized as follows. Section 2 introduces three neural network based multi-channel integration methods. Section 3 discusses the audio-visual multi-channel speech separation networks. The integration of the separation and recognition components is discussed in section 4. Experimental results are presented in section 5. Section 6 draws the conclusions and discuss possible future directions.

\begin{figure*}[ht]
\begin{subfigure}{1\textwidth}
  \centering
  % include first image
  \includegraphics[width=15.5cm]{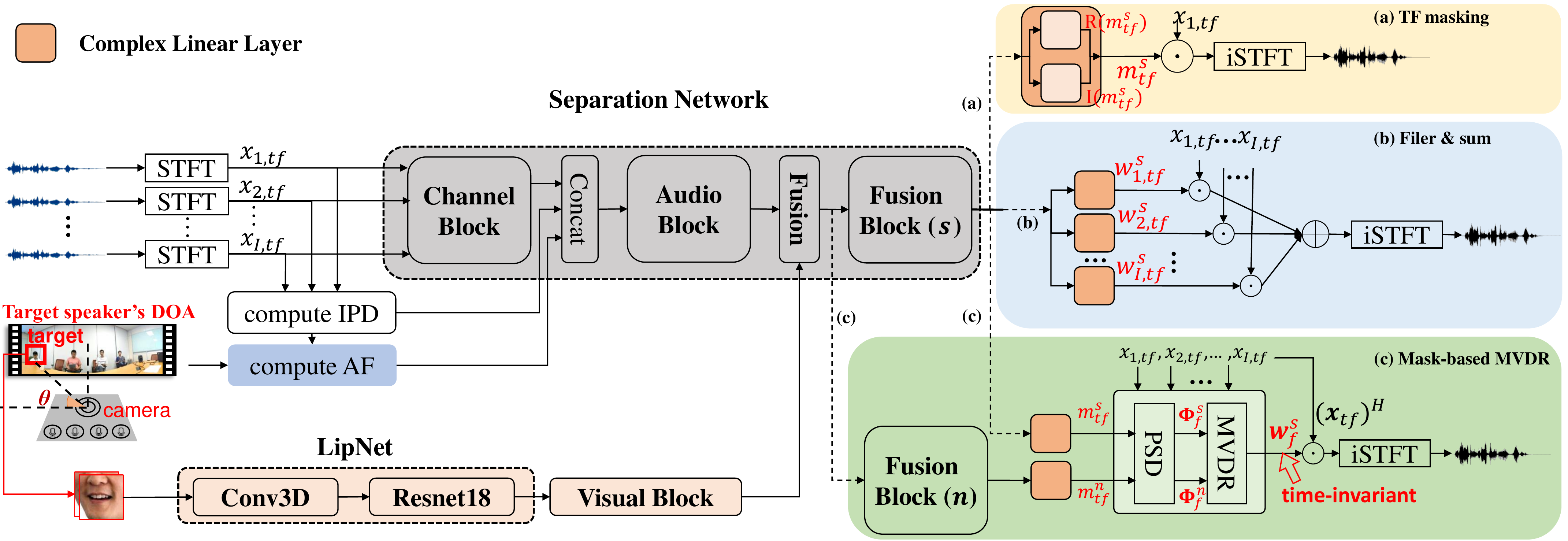}  
%   \caption{TF Masking: $x_{i,tf}$ is the complex spectrum of each channel and $m_{tf}^{s}$ represents the complex mask of the target speaker}
  \label{fig:sub-first}
\end{subfigure}
\vspace{-0.35cm}
\caption{Illustration of the proposed audio-visual multi-channel speech separation networks, where  $x_{i,tf}$ is the complex spectrum of each channel. (a), (b) and (c) represent three options of channel integration approaches: (a) TF masking: $m_{tf}^{s}$ represents the complex mask of the target speaker, where ${\sf R}(m_{tf}^{s}$) and ${\sf I}(m_{tf}^{s}$) are the real and imaginary part of the mask respectively; (b) Filter\&sum: $w^{s}_{i,tf}$ denotes the beamforming filter parameters of the $i$th channel; (c) Mask-based MVDR: $m_{tf}^{s}$ and $m_{tf}^{n}$ are the complex masks of the target and interfering sources, ${\pmb{\Phi}}_f^{s}$ and ${\pmb{\Phi}}_f^{n}$ are the corresponding PSD matrices, ${\pmb{w}}_f^{s}$ is the time-invariant beamforming filter parameters.}
\label{fig:fig}
\end{figure*}

% \begin{figure*}[htb]
% \begin{minipage}[b]{1\linewidth}
%   \centering
%   \centerline{\includegraphics[width=16.5cm]{LaTeX/TFMASK_6.pdf}}
%   \centerline{(a) TF Masking: $x_i$ is the complex spectrum of each microphone array signal and $m^{s}$ represents the complex mask of the targert speaker}\medskip
% %  \vspace{2.0cm}
% \end{minipage}
% \begin{minipage}[b]{0.5\linewidth}
%   \centering
%   \centerline{\includegraphics[width=8cm]{LaTeX/FAS_5.pdf}}
%   \centerline{(b) Filter\&sum: $w^{s}_{i}$ is the beamforming parameters}
% \end{minipage}
% \begin{minipage}[b]{0.5\linewidth}
%   \centering
%   \centerline{\includegraphics[width=8cm]{LaTeX/MVDR_5.pdf}}
%   \captionof{figure}{(c) Mask-based MVDR: $m^{s}$ and $m^{n}$ are the target and interference mask respectively}
% %   \centerline{(c) Mask-based MVDR: $m^{s}$ and $m^{n}$ are the target and interference mask respectively}
% \end{minipage}
% \caption{Illustration of  audio-visual multi-channel speech separation networks.}
% \label{figures}
% \end{figure*}
% \vspace{-0.25cm}
\section{Multi-channel Speech Separation}
This section introduces the three multi-channel speech separation approaches used in this paper, i.e. \textit{TF masking}, \textit{filter\&sum} and \textit{mask-based MVDR}.

% \subsection{\textit{TF masking}}

% \subsection{\textit{Filter\&sum}}

% \subsection{\textit{Mask-based MVDR}}

% The neural network (NN) based \textit{TF masking} approaches \cite{WJTVF, MULTIBAND, CRM} predict spectral time-frequency (TF) masks that specify whether a particular TF bin is dominated by the target speaker or interfering sources to facilitate speech separation.
% The neural \textit{filter\&sum} approaches directly estimate the beamforming filter parameters in either time domain \cite{Taramulti, FASNET1, FASNET2} or frequency domain \cite{DEEPBEAMFORMER} before applying these to channel integration to produce the separated output.
% The more complicated \textit{mask-based MVDR} \cite{IMPMVDR, Angle, ROM, MULTIFAROVER, MIMO2019, MIMO2020} and related \textit{GEV} \cite{GEV2016, Beamnet}  approaches compute the power spectral density (PSD) matrices for the target and overlapping speakers using spectral TF masks to obtain the beamforming parameters.

%%%%%%%%%%%%%%%%%%%%%%%%%%%%%%%%%%%%%%% Section 2 %%%%%%%%%%%%%%%%%%%%%%%%%%%%%%%%%%%%%%%%%%%%%
\vspace{-0.25cm}
\subsection{TF masking}
The \textit{TF masking} approaches predict spectral TF masks that specify whether a particular TF bin is dominated by the target speaker or the interfering sources to facilitate speech separation. 
Previous research has shown that the complex mask (CM) \cite{CRM2015} outperforms the real-value ratio mask (RM) in speech separation and recognition tasks \cite{CRM}.
Therefore, the CM based TF masking approach is adopted in this work.
The complex spectrum of the separated output  $y_{tf}$ is computed as follows:
\begin{equation}
    y_{tf}=m^{s}_{tf}* x_{R,tf},
\end{equation}
where '$*$' indicates complex multiplication, $x_{R,tf}$ is the reference channel's complex spectrum TF bin of the overlapped speech (without loss of generality, we selects $R=1$ in this paper), and $m^{s}_{tf} \in \mathbb{C}$ is the CM of the target speaker. Though the \textit{TF masking} approaches can provide perceptually enhanced sounds, there is a shared belief that the processing artifacts created by the masking are detrimental to the ASR technology \cite{NTTchime3}.
\vspace{-0.25cm}
\subsection{Filter\&sum}
The neural \textit{filter\&sum} approaches directly estimate the beamforming filter parameters in either  time domain \cite{Taramulti, FASNET1, FASNET2} or frequency domain \cite{DEEPBEAMFORMER} in a fully-trainable fashion. In this work, we adopt a frequency domain \textit{filter\&sum} approach to produce the separated outputs:
% The \textit{filter\&sum} approach estimates the beamforming filters in a fully-trainable fashion
\begin{equation}
    y_{tf}=\sum_{i}w_{i,tf} * x_{i,tf}.
\end{equation}
Where $w_{i,tf}$ is the complex value beamforming parameters corresponding to the $i$th channel.
% The neural \textit{filter\&sum} approaches directly estimate the beamforming filter parameters in either time domain \cite{Taramulti, FASNET1, FASNET2} or frequency domain \cite{DEEPBEAMFORMER} before applying these to channel integration to produce the separated output.
%  This approach estimate the beamforming filters in a fully-trainable fashion: 
% \begin{equation}
%     y_{tf}=\sum_{i}w_{i,tf}\odot x_{i,tf}
% \end{equation}
% where $w_{i,tf}$ is the mask of the $i$th channel.
\vspace{-0.25cm}
\subsection{Mask-based MVDR}
% The more complicated \textit{mask-based MVDR} \cite{IMPMVDR, Angle, ROM, MULTIFAROVER, MIMO2019, MIMO2020} and related \textit{GEV} \cite{GEV2016, Beamnet}  approaches compute the power spectral density (PSD) matrices for the target and overlapping speakers using spectral TF masks to obtain the beamforming parameters.
The more complicated \textit{mask-based MVDR} beamforming approach \cite{IMPMVDR, MultiMMSE} has demonstrated state-of-the-art performance in noisy and overlapped speech recognition  \cite{ROM, MULTIFAROVER, MIMO2019}. Such approach first uses deep neural networks to estimate the real-value\cite{ROM, MULTIFAROVER, MIMO2019} or complex\cite{xu2020neural} TF mask of the target speech $m^s_{tf}$ and other interfering sources $m^n_{tf}$ respectively. The PSD matrices corresponding to each source are then calculated as follows:
% estimated mask of each source are used to compute the PSD matrices $\Phi_{f}^{j}$ for $j\in{s,n}$: 
\begin{align}
    {\pmb{\Phi}}_{f}^{s}=\frac{1}{\sum_{t=1}^{T} m^{s}_{tf}*(m^{s}_{tf})^{\sf H}}\sum_{t=1}^{T} (m^{s}_{tf}*{\pmb{x}}_{tf})({m^{s}_{tf}*\pmb{x}}_{tf})^{\sf H}, \nonumber\\
    {\pmb{\Phi}}_{f}^{n}=\frac{1}{\sum_{t=1}^{T} m^{n}_{tf}*(m^{n}_{tf})^{\sf H}}\sum_{t=1}^{T} (m^{n}_{tf}*{\pmb{x}}_{tf})({m^{n}_{tf}*\pmb{x}}_{tf})^{\sf H},
\end{align}
where $(\cdot)^{\sf H}$ denotes the conjugate transpose,  ${\pmb{x}}_{tf}=[x_{1,tf},..,x_{I,tf})\in\mathbb{C}^I$ is a complex vector containing the TF bins of all $I$ microphone array channels. ${\pmb{\Phi}}_{f}^{s}$ and ${\pmb{\Phi}}_{f}^{n}$ represent the PSD matrices of the target and other interfering sources respectively. The time-invariant beamforming filter parameters ${\pmb{w}}_{f}^{s}$ of the target speech are then obtained by the solution of MVDR beamformer as:
\begin{equation}
    {\pmb{w}}_{f}^{s}=\frac{( {\pmb{\Phi}}_{f}^{n})^{-1} {\pmb{\Phi}}_{f}^{s}}{{\text{Trace}}(( {\pmb{\Phi}}_{f}^{n})^{-1} {\pmb{\Phi}}_{f}^{s})}{\pmb{u}},
\end{equation}
where ${\pmb{u}}=[1,0,...,0]^T$. Finally, the beamforming filters ${\pmb{w}}_{f}^{s}$ are used to compute the separated spectrum $y_{tf}$ as follows:
\begin{equation}
    y_{tf}=({\pmb{w}}_{f}^{s})^{\sf H}{\pmb{x}}_{tf}.
\end{equation}

%%%%%%%%%%%%%%%%%%%%%%%%%%%%%%%%%%%%%%%%%%%% Section 3 %%%%%%%%%%%%%%%%%%%%%%%%%%%%%%%%%%%%%%%%%%%%%%%

\section{Audio-visual Multi-channel Separation}
This section presents our audio-visual multi-channel speech separation networks.

\noindent{\bf{Audio inputs:}} As shown in Figure 1, the complex spectrum of all the microphone array channels are first computed through short-time Fourier transform (STFT). 
The inter-microphone phase differences (IPDs) \cite{ROM}, which reflect the time difference of arrival (TDOA), are also used as input features:
% \begin{equation}
%     {\rm{IPD}}^{(i,j)}_{tf}=\angle(\frac{x_{i,tf}}{x_{j,tf}}),
% \end{equation}
\begin{equation}
    {\rm{IPD}}^{(i,j)}_{tf}=\angle({x_{i,tf}}/{x_{j,tf}}),
\end{equation}
\noindent where $x_{i,tf}$ represents the i-th channel's complex spectrum of the mixed signal at time frame $t$ and frequency bin $f$, and $\angle (\cdot)$ outputs the angle of the input pair of channel specific TF spectrum. 
Given the direction of arrival (DOA) of the target speaker, e.g. by tracking the speaker's face from a 180-degree wide-angle camera as shown in Figure 1, a location-guided angle feature (AF) introduced in \cite{Angle,gu2020multi} is adopted to provide the target discriminative information:
% \begin{equation}
%     {\rm{AF}}_{\theta,tf}=\sum_{m=1}^{M}\frac{e^{(i,j)}_{\theta,f}\frac{x_{i,tf}}{x_{j,tf}}}{|e^{(i,j)}_{\theta,f}\frac{x_{i,tf}}{x_{j,tf}}|},
% \end{equation}
% where $e^{(i,j)}_{\theta,f}$ is the steering vector coefficients for target speaker from direction $\theta$ at frequency $f$ corresponding to microphone pair $(i, j)$ and $M$ is the number of selected microphone pairs.
\begin{align}
    &{\rm{AF}}_{\theta,tf}=\sum_{m=1}^{M} \langle{\sf \pmb{e}}^{{\rm{pd}}^{(i,j)}_{\theta, tf}}, {\sf \pmb{e}}^{{\rm{IPD}}^{(i,j)}_{tf}} \rangle, \nonumber \\ 
    & {\rm{pd}}^{(i,j)}_{\theta, tf} = 2\pi f f_{s} d_{ij}\cos(\theta)/(2(F-1)c),
\end{align}
where $\pmb{e}^{(\cdot)} =[\cos(\cdot), \sin(\cdot)]$, $M$ is the number of selected microphone pairs, ${\rm{pd}}^{(i,j)}_{tf}$ represents the phase delay between $i$th and $j$th microphone of a plane wave from direction $\theta$, $d_{ij}$ is the distance between $i$th and $j$th microphone, c is the sound velocity, $f_s$ is the sample rate and $F$ is the number of TF bins.
% Intuitively, the angle feature indicates whether a time-frequency (T-F) bin of the mixture is dominated by the sound from the desired direction, which informs the network to attend to the target speaker. 
% In practice, the arrival direction of the target speaker can be obtained by tracking the speaker's face from a 180-degree wide-angle camera, as shown in Figure 1 (a).
% In practice, the arrival direction of the target speaker can be obtained by tracking the speaker's face from a video captured by a 180-degree wide-angle camera, as shown in Figure 1 (a).

\noindent{\bf{Visual inputs:}} 
The visual inputs are shown in the pink part of Figure 1.
Considering that the visual modality is invariant to acoustic corruption, this paper leverages the visual modality containing speaker-specific information to improve the  estimation of masks or filter parameters.
In this work, a LipNet consisting of a 3D convolutional layer and a 18-layer ResNet is used to extract the lip embeddings from the lip region of the target speaker. 
Such LipNet is first trained on a lipreading task as described in \cite{LIPNET}.
The lip embeddings extracted by the LipNet are sent into the visual block before being fused with audio modality.

\noindent{\bf{Modality fusion:}} 
In this work, we adopt a factorized attention-based modality fusion method proposed in our previous work \cite{gu2020multi}, which has been proven to outperform the concatenation method.
This method firstly factorizes the mixed audio into a set of acoustic subspaces, then leverages the target's information from the visual modality to enhance these subspace acoustic embeddings with learnable weights. Please refer to our previous work \cite{gu2020multi} for details.

The outputs of the fusion layer are sent into the fusion blocks to compute the CM masks or beamforming filter parameters. Figure 1 (a) shows the diagram of the \textit{TF masking} approach, which estimates the CM mask $m^{s}_{tf}$ of the target speaker. The diagram of the \textit{filter\&sum} approach is shown in Figure 1 (b), which estimates the beamforming parameters $w_{i,tf}$  of several microphone array channels. The \textit{mask-based MVDR} approach estimates the masks of the target and interfering sources simultaneously before feeding into a MVDR solution layer implementing Eq.(4) and (5), as shown in Figure 1 (c). 
The Si-SNR loss function is used to train the separation networks.
Since dereverberation is beyond the scope of this paper, the reverberant non-overlapped speech signal is used as the supervision, following \cite{gu2020multi,FASNET1}.

\vspace{-0.25cm}
\section{Integration of Separation \& Recognition}
\begin{figure}[b]
\begin{minipage}[b]{1\linewidth}
\vspace{-0.7cm}
  \centering
  \centerline{\includegraphics[width=8cm]{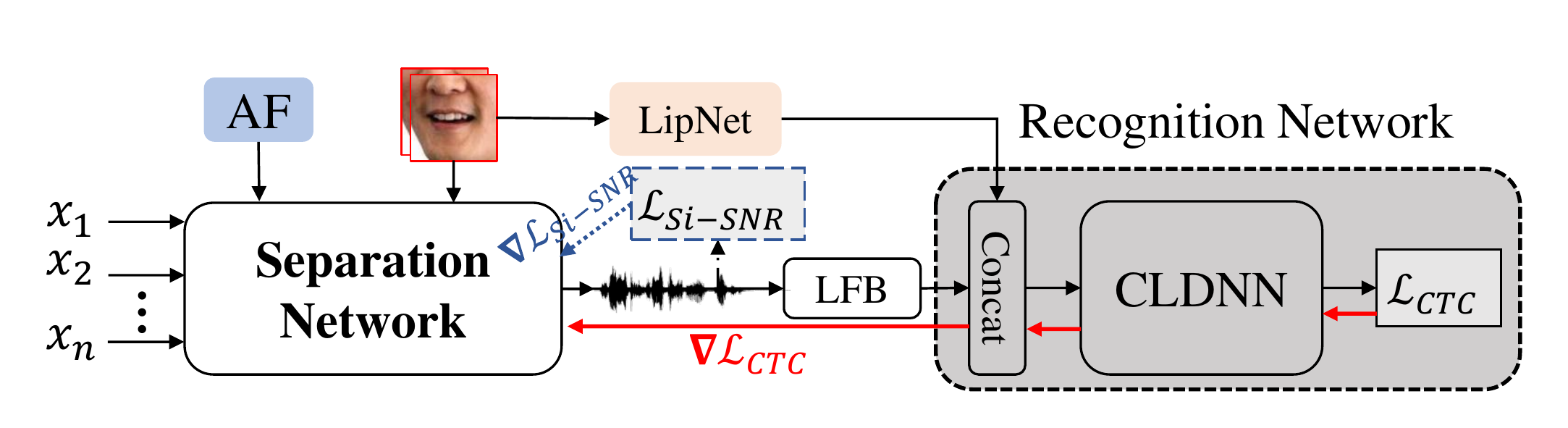}}
\end{minipage}
\caption{Joint fine-tuning: $\nabla\mathcal{L}_{CTC}$ and $\nabla\mathcal{L}_{Si-SNR}$ represent the gradients of CTC and SI-SNR loss functions respectively, "LFB" denotes log filter bank acoustic features.}
\label{figures}
\end{figure} 
Traditionally, the speech separation and recognition components are developed separately and then used in a pipelined fashion \cite{IMPMVDR, ROM, Angle, MULTIFAROVER}. 
However, two issues arise with such approach: 
1) the cost function mismatch between separation and recognition components cannot guarantee the separated outputs target to optimal recognition performance; 
2) the artifacts created by separation can increase modeling confusion of the recognition component and lead to performance degradation. 
% Although speech separation networks can be performed by combining independently trained speech recognition systems \cite{IMPMVDR, ROM, Angle, MULTIFAROVER}, the artifacts created by the separation networks unknown to the recognition system can disturb the whole systems' performance. 
% In addition, the error cost mismatch between the separation and recognition component can lead the system to run into sub-optimal. 
According to \cite{Beamnet,CRM, JOINT, MAX2019}, tight integration of the two components with joint ﬁne-tuning can address above two issues.
% According to \cite{Beamnet,CRM, JOINT}, using a tightly integration these issues can be mitigated by a tightly integration of the two components using jointly fine-funing.

% \subsection{Recognition network}
\noindent {\bf{Recognition network:}} 
The architecture of our audio-visual speech recognition (AVSR) network is shown in Figure 2. 
The lip embeddings extracted from the LipNet is concatenated with the log filter bank acoustic features extracted from the separated waveform. 
The concatenated features are sent into the convolutional long short-term memory deep neural networks (CLDNN) to generate the frame level mono-phone posteriors. 
The recognition network is optimized using the CTC loss function.

% \subsection{Integration of separation \& recognition}
\noindent {\bf{Integration of separation \& recognition:}} To tightly integrate the separation and recognition components, here we investigate three variants of fine-tuning methods: 
% 1) fine-tuning the recognition system only on the enhanced signals as in the pipelined approach;
1) fine-tuning the recognition system only on the enhanced signals;
2) jointly fine-tuning the separation and recognition components using the CTC cost function; 3) jointly fine-tuning both systems using a multi-task criterion, which interpolates the  CTC and Si-SNR cost function:
\begin{equation}
    \mathcal{L} = \mathcal{L}_{CTC} + \alpha \mathcal{L}_{Si-SNR},
\end{equation}
where $\alpha$ is a manually tuned weight of the Si-SNR loss.

\vspace{-0.25cm}
\section{Experiment \& Results}
% This section first presents our experiment setup and then shows the experiment results.

\subsection{Experiment Setup}
\vspace{-0.1cm}
% \subsection{Simulated overlapped speech}
\noindent{\bf{Simulated overlapped speech:}} 
The multi-channel overlapped speech used in the system development is simulated using the LRS2 dataset \cite{AVSR1}. 
The simulated dataset is split into three subsets with 12.5k, 4.6k and 1.2k utterances respectively for training, validation and evaluation. 
% The simulated overlapped speech contains 12.5k, 4.6k and 1.2k utterances for training, validation and testing. 
The details of the simulation process is described in \cite{gu2020multi}.
% We use a 15-element non-uniform linear array with spacing 7-6-5-4-3-2-1-1-2-3-4-5-6-7 cm. 
% The overlapped multi-channel audio signals are generated by convolving single channel signals with Room Impluse Respones (RIRs) simulated by image-source method. 
% The room size is ranging from 4m-4m-2.5m to 10m-8m-6m (length-width-height). 
% The speakers and the microphone array randomly located in the room at least 0.3m away from the wall. 
% The distance between the speaker and microphones ranges from 1m to 5m. 
% The reverberation time T60 is sampled in a range of 0.05s to 0.7s. The signal-to-interference ratio (SIR) is ranging from -6 to 6dB.  
% Also, noise with 18-30 dB SNR is added to all the multi-channel speech mixtures. The angles between the target and interfering speakers are ranging from $0^\circ$ to $180^\circ$. 

% \subsection{Replayed overlapped speech}
\noindent{\bf{Replayed overlapped speech:}}
To further evaluate the systems' performance, 1.2k overlapped speech utterances are recorded in a meeting room of the size 10m$\times$5m$\times$3m.
To generate overlapped speech, two loudspeakers are used to replay different sentences of the LRS2 test set simultaneously.  
The structure of the microphone array used during recording is the same as that used in simulation.
% The microphone array used in recording shares the same structure as the array used in the simulation process.
The target and interfering speakers are located at following directions 
(15$^\circ$,30$^\circ$), (45$^\circ$,30$^\circ$), (75$^\circ$,30$^\circ$), (105$^\circ$,30$^\circ$), (30$^\circ$,60$^\circ$), (90$^\circ$,60$^\circ$), (120$^\circ$,60$^\circ$), (150$^\circ$,60$^\circ$) 
and the distance between the loudspeakers and microphones ranges from 1m to 1.5m.
The average overlapping ratio of the replayed overlapped speech is around 80\% and SIR is around 1.5dB.

\noindent{\bf{Model architecture:}}
1) 
% {\color{red}{In the separation networks, the channel, audio and fusion blocks are developed using 1, 1 and 3 repeats of 8 convolutional blocks \cite{TCN} with the set of $B=256, H=512, R=3$ as described in \cite{gu2020multi}. The visual block consists of 5 convolutional blocks. Batch normalization is adopted in each block.}} 
The audio-visual multi-channel separation network is developed based on the time convolutional network (TCN) structure \cite{TCN}. More details can be found in our previous paper \cite{gu2020multi}. 
In the \textit{filter\&sum} approach, a series of complex linear layers are used to estimate the filter parameters of each channel using the fusion block's outputs. 
For all \textit{filter\&sum} systems in the remaining part of this paper \footnote{\noindent Adding more channels up to 15 microphones produces limited improvement for \textit{filter\&sum} systems, while increasing the computational cost.}, filter parameters are estimated using only the first and eighth channels.
% In preliminary experiment, we found that estimating the filter parameters of all 15 microphones is computationally heavy and show limited improvement. In contrast, estimating filter parameters of channel $(0, 7)$ provides acceptable performance with reasonable computational costs. Therefore, in this paper, we only report the results of \textit{filter\&sum} approach estimating filter parameters of channel $(0, 7)$. 
% {\color{red}{In the \textit{mask-based mvdr} approach, an additional fusion block is added to compute the mask of the interfering sources.}} 
2) The recognition network starts with four 2-dimensional convolutional layers with channel sizes (64, 64, 128, 128) and  kernel size $3{\rm{x}}3$ followed by four 1280 hidden units BLSTM layers and one softmax layer. The language model (LM) used in recognition is a 4-gram LM developed on 2.33M words of transcripts of LRS2 training and pretrain set.
3) In multi-task fine-tuning Eq.(8), $\alpha$ is set to 0.1 for \textit{TF masking} approach and  1 for \textit{filter\&sum} and \textit{mask-based mvdr} approaches.

\noindent{\bf{Features:}} 
1) For the separation networks, 257-dimensional complex spectrum are used, which are extracted with a 32ms window and 16ms frame rate. The AF and IPDs are extracted between 9 microphone pairs (1,15), (2, 14), (3, 13), (1, 7), (12, 4), (11, 5), (12, 8), (7, 10), (8, 9). These pairs are selected to sample different spacing between microphones \cite{MULTIBAND, gu2020multi}. The ground truth direction $\theta$ of the target speaker is used during training and evaluation.
% During training and evaluation, the ground truth target speaker’s direction is used.
% During training and simulated evaluation, the ground truth target speaker’s direction is used. For the replayed evaluation, an approximate direction is used; 
2) For the recognition network, 40-dimensional log filter bank features are used, which are extracted using a 40ms window and 10ms frame rate.
3) For visual inputs, we crop the already centered visual frames to 112 by 112 pixels and up-sample them to align with the audio frames via linear interpolation. 
\vspace{-0.25cm}
\subsection{Recognition results on non-overlapped speech}
Table 1 presents the WER results of our CLDNN based ASR and AVSR systems on non-overlapped speech in anechoic and simulated reverberant environments. 
Since we are not aiming for dereverberation in our overlapped speech recognition systems, the WER on reverberant non-overlapped speech (last line) can be viewed as an upper bound for all subsequent experiments on overlapped speech. 
\vspace{-0.3cm}
\begin{table}[htb]
    \caption{Performance of ASR and AVSR systems on echo free and reverberant non-overlapped speech.}
    \label{tab:my_label}
    \footnotesize
    \centering
    {\scalebox{0.9}{
    \begin{tabular}{l|c|c}
    \toprule
    \multicolumn{1}{c|}{Data} & System &WER (\%) \\
    \hline
    \multirow{2}{*}{Anechoic non-overlapped speech} & ASR  &11.04 \\
     & AVSR  &{\bf{9.77}} \\
    \hline
    \hline
    \multirow{2}{*}{Reverberant non-overlapped speech}& ASR  &15.33 \\
    & AVSR  &{\bf{13.93}} \\
    \bottomrule
    \end{tabular}}
    }
\end{table}
\vspace{-0.65cm}

\subsection{Audio-only vs. audio-visual systems}
The performance of the audio-only and audio-visual overlapped speech recognition systems trained using simulated overlapped speech is shown in Table 2. All the multi-channel systems (line 3-10) are jointly fine-tuned using the CTC loss function. 
Several trends can be observed in Table 2. 
1) The first block (line 1--2) in Table 2 presents the recognition performance of monaural ASR and AVSR systems without using microphone array and explicit speech separation components. For these very simple systems, simply adding the visual modality in the recognition network can approximately halve the WER, which confirms the findings in our previous research \cite{AVSRJIAN}.
2) The second block (line 3--6) of Table 2 shows the results of the multi-channel audio-only systems. "Delay \& Sum" means applying frequency domain delay and sum beamforming approach using the steering vector computed by the given array structure and ground truth DOA, as described in \cite{DSB}. 
Compared with the monaural ASR system (line 1), the multi-channel speech separation components can significantly improve the systems' performance by up to 49.98\% (comparing line 1 \& 6 on simu) absolute WER reduction. 
However, there is a large performance gap (around 13\%) between simulated and replayed data using NN based multi-channel audio-only systems (line 4--6).
3) The performance of the proposed audio-visual multi-channel overlapped speech recognition systems is shown in the third block (line 7--10) of Table 2. Comparing the results of audio-visual and audio-only multi-channel systems, it can be seen that leveraging visual modality in both separation and recognition components can reduce the WER ranging from 6.81\% (comparing line 6 \& 10 on simu) to 28.72\% (comparing line 4 \& 8 on replay). Moreover, the performance gap between the simulated and replayed overlapped speech is much smaller compared to that on the audio-only systems, which suggests that the proposed audio-visual multi-channel speech recognition systems are more robust.  

% 2) The second block of Table 2 shows the results of the audio-only systems, where oracle delay and sum means applying traditional filter and sum beamforming technique on frequency domain using a given direction-of-arrival. 
% Compared with the ASR systems without any speech separation components, the multi-channel speech separation techniques can greatly improve the recognition performance.
% It can be observed that the \textit{mask-based MVDR} approach achieve the best performance of audio-only overlapped speech recognition on both simulated and replayed test data. 
% However, there is a large performance gap between simulated and replayed data using audio-only neural network based methods.
% However, there is a large performance gap of neural network based methods between simulated and replayed data. 
% 3) The performance of the proposed audio-visual multi-channel overlapped speech recognition system is shown in the third block of Table 2. 
% Comparing the results of audio-visual and audio-only multi-channel systems, it can be seen that leveraging visual modality in both separation and recognition components can greatly decrease the WER. 
% The \textit{filter\&sum} and \textit{mask-based MVDR} approaches achieve the best performance on the simulated and replayed data respectively.
% The performance gap between the simulated and replayed overlapped speech is much smaller than the audio-only system, which indicates that the proposed audio-visual multi-channel speech recognition systems are more robust. 

\vspace{-0.25cm}
\begin{table}[htb]
    \footnotesize
    \caption{Performance of audio-only and audio-visual overlapped speech recognition systems using various channel integration methods. The separation and recognition components are jointly fine-tuned using the CTC loss. "AF" denotes angle feature, "raw" denotes raw signal of the first channel.}
    \vspace{-0.25cm}
    \label{tab:my_label}
    \centering
    {\scalebox{0.85}{
    \begin{tabular}{c|l|c|c|c|c|c}
    \toprule
    &\multicolumn{3}{c|}{Separation} & Recognition & \multicolumn{2}{c}{WER(\%)} \\
    \cline{2-7}
    & \multicolumn{1}{c|}{method}  & AF & +visual  &  +visual & simu & replay\\
    \hline
    \hline
    1&\multicolumn{3}{c|}{raw}              & \xmark & 75.36 & 80.55 \\
    2&\multicolumn{3}{c|}{raw}              & \cmark & 32.06 & 31.93\\
    \hline
    \hline
    3&Delay \& sum & \cmark                         & \xmark & \xmark & 49.25 & 44.34\\
    4&TF masking    & \cmark                         & \xmark & \xmark & 33.12 & 46.75\\
    5&Filter \& Sum   & \cmark    & \xmark & \xmark & 30.24 & 43.83\\
    6&Mask-based MVDR   & \cmark                     & \xmark & \xmark & \bf{25.38} & \bf{39.07}\\
                                        %   & (0, 2, 4, 7) & \xmark & ASR & 33.59 \\
                                        %   & (0, 2, ... , 14) & \xmark & ASR & 32.47 \\
    \hline
    \hline
    7&Delay \& Sum      & \cmark                    & \xmark & \cmark & 25.81 & 24.46\\
    8&TF masking        & \cmark                     & \cmark & \cmark & 19.25 & {18.03}\\
    9&Filter \& Sum     & \cmark   & \cmark & \cmark & {\bf{17.21}} & 19.87\\
    10&Mask-based MVDR  & \cmark                      & \cmark & \cmark & 18.57 & \bf{16.85}\\
                                        %   & (0, 2, 4, 7) & \cmark & AVSR & 18.51\\
                                        %   & (0, 2, ... , 14) & \cmark & AVSR & 20.83\\
    \bottomrule
    \end{tabular}}
    }
\end{table}
\vspace{-0.65cm}

\subsection{Comparison of different fine-tuning approaches}
The results of different fine-tuning approaches are listed in Table 3. 
The first line shows the baseline systems using the CTC loss function to fine-tune the recognition components only while keeping the parameters of the separation components fixed. 
Jointly fine-tuning the separation and recognition components using the CTC loss function (line 2) can improve the systems' performance by 0.4\% to 5.2\% WER reduction. The best results are obtained using a multi-task interpolation between the CTC and Si-SNR cost function to fine-tune the entire systems (last line).

% Systems fine-tuning both the separation and recognition components using the CTC loss function, which outperform the baseline systems by 
% Comparing with the baseline systems using the CTC loss function only on the recognition components while keeping the separation components fixed, applying the CTC loss to fine-tune both the separation and recognition components can improve the systems' performance from 0.4\% to 5.2\% WERR . 

% Several trends can be observed in this table:
% % We can observe that fine-tuning the recognition system using the separated outputs can improve the system performance by approximate 10\% WERR. 
% We can observe that finetuing the recognition network with the CTC loss only while keeping the separation component and its outputs can improve the system performance by approximate 10\% WERR.
% Jointly fine-tuning the separation and recognition components using CTC cost can further improve the systems' performance. 
% However, the improvement of using the interpolation of CTC and Si-SNR loss is limited against using only CTC loss function.
\vspace{-0.25cm}
\begin{table}[htb]
    \caption{Performance of different fine-tuning approaches of audio-visual multi-channel speech recognition systems.}
    \label{tab:my_label}
    \footnotesize
    \centering
    \vspace{-0.25cm}
    {\scalebox{0.8}{
    \begin{tabular}{ccccccccccc}
    \toprule
    \multicolumn{3}{c}{Fine-tuning} & {TF masking} &{Filter\&sum} &{MVDR}\\
    \cline{0-2}
    Sep. & Recg.        & Loss                    & simu/replay & simu/replay & simu/replay\\
    \hline
    % \xmark     & \xmark             & NA   & 33.2/33.4 & 29.7/36.4 & 33.2/25.7 \\
    \xmark     & \cmark             & $\mathcal{L}_{CTC}$   & 22.9/23.2 & 19.2/24.1 & 19.3/17.3 \\
    \cmark     & \cmark             & $\mathcal{L}_{CTC}$   & 19.3/18.0 & 17.2/19.9 & 18.6/{\bf{16.9}} \\
    \cmark     & \cmark             & $\mathcal{L}_{CTC}+\alpha\mathcal{L}_{Si-SNR}$   & {\bf{18.6}}/{\bf{18.0}} & {\bf{16.1}}/{\bf{19.2}} & {\bf{18.4}}/{\bf{16.9}} \\
    
    \bottomrule
    \end{tabular}}
    }
\end{table}
\vspace{-0.65cm}

\section{Conclusions \& Future Work}
This paper presents an audio-visual multi-channel overlapped speech recognition system with tightly integrated separation front-end and recognition back-end. 
Three multi-channel integration approaches, i.e. \textit{TF masking}, \textit{filter\&sum} and \textit{mask-based MVDR} are investigated in the system development. 
The experiment results suggest that: 
1) using visual modality can improve the systems' performance and robustness; 
2) jointly fine-tuning the separation and recognition components can tightly integrate the two components for better speech recognition performance. 
In the future, this work will be extended to: 
1) performing separation and dereverberation simultaneously in the separation front-end; 
2) applying to more challenging applications, such as the situation when both the visual and audio  are degraded;
3) investigating other separation and recognition architectures.

\section{Acknowledgements}
The authors would like to thank Shansong Liu for the insightful discussion. This research is supported by Hong Kong Research Grants Council General Research Fund No.14200218, Theme Based Research Scheme T45-407/19N  and Shun Hing Institute of Advanced Engineering Project No. MMT-p1-19.

\newpage

\bibliographystyle{IEEEtran}

\bibliography{mybib}

% \begin{thebibliography}{9}
% \bibitem[1]{Davis80-COP}
%   S.\ B.\ Davis and P.\ Mermelstein,
%   ``Comparison of parametric representation for monosyllabic word recognition in continuously spoken sentences,''
%   \textit{IEEE Transactions on Acoustics, Speech and Signal Processing}, vol.~28, no.~4, pp.~357--366, 1980.
% \bibitem[2]{Rabiner89-ATO}
%   L.\ R.\ Rabiner,
%   ``A tutorial on hidden Markov models and selected applications in speech recognition,''
%   \textit{Proceedings of the IEEE}, vol.~77, no.~2, pp.~257-286, 1989.
% \bibitem[3]{Hastie09-TEO}
%   T.\ Hastie, R.\ Tibshirani, and J.\ Friedman,
%   \textit{The Elements of Statistical Learning -- Data Mining, Inference, and Prediction}.
%   New York: Springer, 2009.
% \bibitem[4]{YourName17-XXX}
%   F.\ Lastname1, F.\ Lastname2, and F.\ Lastname3,
%   ``Title of your INTERSPEECH 2020 publication,''
%   in \textit{Interspeech 2020 -- 20\textsuperscript{th} Annual Conference of the International Speech Communication Association, September 15-19, Graz, Austria, Proceedings, Proceedings}, 2020, pp.~100--104.
% \end{thebibliography}

\end{document}